\documentclass[english,prl,aps,floatfix,twocolumn,longbibliography,superscriptaddress]{revtex4-1}
\usepackage[T1]{fontenc}
\usepackage[latin9]{inputenc}
\usepackage{color}
\usepackage{verbatim}
\usepackage{amsmath}
\usepackage{amssymb}
\usepackage{graphicx}
\usepackage{wasysym}

\makeatletter

\usepackage[colorlinks=true,linkcolor=blue,anchorcolor=red,citecolor=blue,urlcolor=blue]{hyperref}
\usepackage{fancyhdr}

\makeatother

\usepackage{babel}
\begin{document}
\renewcommand{\figurename}{Fig.}
\title{Super-resonant transport of topological surface states subjected to
in-plane magnetic fields}
\author{Song-Bo Zhang}
\email{songbo.zhang@physik.uni-wuerzburg.de}

\affiliation{Institut für Theoretische Physik und Astrophysik, Universit\"at W\"urzburg,
97074 W\"urzburg, Germany}
\author{Chang-An Li}
\affiliation{Institut für Theoretische Physik und Astrophysik, Universit\"at W\"urzburg,
97074 W\"urzburg, Germany}
\author{Francisco Pe\~na-Benitez}
\affiliation{Max-Planck-Institut f\"ur Physik komplexer Systeme, N\"othnitzer
Strasse 38, 01187 Dresden, Germany}
\affiliation{W\"urzburg-Dresden Cluster of Excellence ct.qmat, Germany}
\author{Piotr Sur\'owka}
\affiliation{Max-Planck-Institut f\"ur Physik komplexer Systeme, N\"othnitzer
Strasse 38, 01187 Dresden, Germany}
\affiliation{W\"urzburg-Dresden Cluster of Excellence ct.qmat, Germany}
\affiliation{Department of Theoretical Physics, Wroc\l{}aw University of Science
and Technology, 50-370 Wroc\l{}aw, Poland}
\author{Roderich Moessner}
\affiliation{Max-Planck-Institut f\"ur Physik komplexer Systeme, N\"othnitzer
Strasse 38, 01187 Dresden, Germany}
\affiliation{W\"urzburg-Dresden Cluster of Excellence ct.qmat, Germany}
\author{Laurens W. Molenkamp}
\affiliation{Physikalisches Institut (EP3), Universit\"at W\"urzburg, Am Hubland,
97074 Würzburg, Germany}
\affiliation{Institute for Topological Insulators, Universit\"at W\"urzburg,
Am Hubland, 97074 Würzburg, Germany}
\affiliation{W\"urzburg-Dresden Cluster of Excellence ct.qmat, Germany}
\affiliation{Max Planck Institute for Chemical Physics of Solids, Dresden D-01187,
Germany}
\author{Bj\"orn Trauzettel}
\affiliation{Institut für Theoretische Physik und Astrophysik, Universit\"at W\"urzburg,
97074 W\"urzburg, Germany}
\affiliation{W\"urzburg-Dresden Cluster of Excellence ct.qmat, Germany}
\date{\today}
\begin{abstract}
Magnetic oscillations of Dirac surface states of topological insulators are typically expected to be associated with the formation of Landau levels or the Aharonov-Bohm effect. We instead study the conductance of Dirac surface states subjected to an in-plane magnetic field in presence of a barrier potential. Strikingly, we find that, in the case of large barrier potentials, the surface states exhibit pronounced oscillations in the conductance when varying the magnetic field, in the \textit{absence} of Landau levels or the Aharonov-Bohm effect. These novel magnetic oscillations are attributed to the emergence of \textit{super-resonant transport} by tuning the magnetic field, in which many propagating modes cross the barrier with perfect transmission. In case of small and moderate barrier potentials, we identify a positive magnetoconductance due to the increase of the Fermi surface by tilting the surface Dirac cone. Moreover, we show that for weak magnetic fields, the conductance displays a shifted sinusoidal dependence on the field direction with period $\pi$ and phase shift determined by the tilting direction with respect to the field direction. Our predictions can be applied to various topological insulators, such as HgTe and Bi$_{2}$Se$_{3}$, and provide important insights into exploring and understanding exotic magnetotransport properties of topological surface states.
\end{abstract}

\maketitle
\textit{Introduction.}\textemdash Topological
insulators host gapless surface states which stem from nontrivial bulk topology \citep{HasanKane2010,QiXL11RMP,Shen12book}. These surface states can be modeled by a single Dirac cone. Over the last two decades, topological insulators have been discovered in numerous materials \citep{Fuliang07PRB,Sato10prl,Hasan11NC,Tanaka12nphys, SYXu12ncomms, Hsieh12ncomms,Dziawa12nmater,Yan12rpp,Ando13jpsj}
including HgTe \citep{Brune2011prl}, Bi$_{1-x}$Sb$_{x}$ \citep{Hsieh08Natphy}
and Bi$_{2}$Se$_{3}$ \citep{ZhangHJ_09nphys,Chen09sci,Hsieh09prl,Xia09natphys}.
Magnetotransport of Dirac surface states has been an active research topic  \citep{Taskin09prb,Qu10sci,Analytis2010NatPhys,Chengpeng2010prl, ZRen10prb,Taskin12prl,Chen10prl,Lu11prl, He11prl,Tkachov11prb,Garate12prb, Bardarson13rpp,Sitte2012prl,XLiWang12prl, Peng10natmat,ZhangY10prl, Bardarson10prl,YXu14NatPhys,Yoshimi15NC, FuYishuang_2014nphys,Ilan15PRL, ZhangSB15srep,YXu16NC,Tu16NC, Burset15PRB,Taskin17nc,Dziom17NC, PHe18nphys,SBZhang19PRL,Assouline19NC, HWu19prl,SDZheng20prb,Dyrda20prl}, theoretically and experimentally, since the discovery of topological insulators. It provides vital features, which include particularly magnetic oscillations, to detect and characterize Dirac surface states. Magnetic oscillations are usually associated with the formation of Landau levels or the Aharonov-Bohm effect \citep{Taskin09prb,Qu10sci,Analytis2010NatPhys, Chengpeng2010prl, ZRen10prb,Taskin12prl,Peng10natmat,ZhangY10prl, Bardarson10prl,YXu14NatPhys,Yoshimi15NC,FuYishuang_2014nphys}. Thus, a fundamentally intriguing question is whether magnetic oscillations of topological surface states can appear in the absence of Landau levels or the Aharonov-Bohm effect. In slab geometries of topological insulators, in-plane magnetic oscillations as a function of field strength are often times observed \citep{Sulaev15NL,Taskin17nc,Gracy-unpunblish}. To the best of our knowledge, a convincing explanation of these oscillations is still lacking.

Notably, in typical topological insulators, electron-hole symmetry in the energy spectrum of surface states is broken by the presence of higher-order momentum corrections \citep{Shan11njp,CXLiu10prb,Jost17PNAS}. To fully understand the transport properties of surface states in realistic systems, the consideration of this electron-hole asymmetry is important. Interestingly, the interplay of electron-hole asymmetry and in-plane magnetic fields tilts the surface states at low energies
\citep{SDZheng20prb}.

\begin{figure}[t]
\includegraphics[width=1\columnwidth]{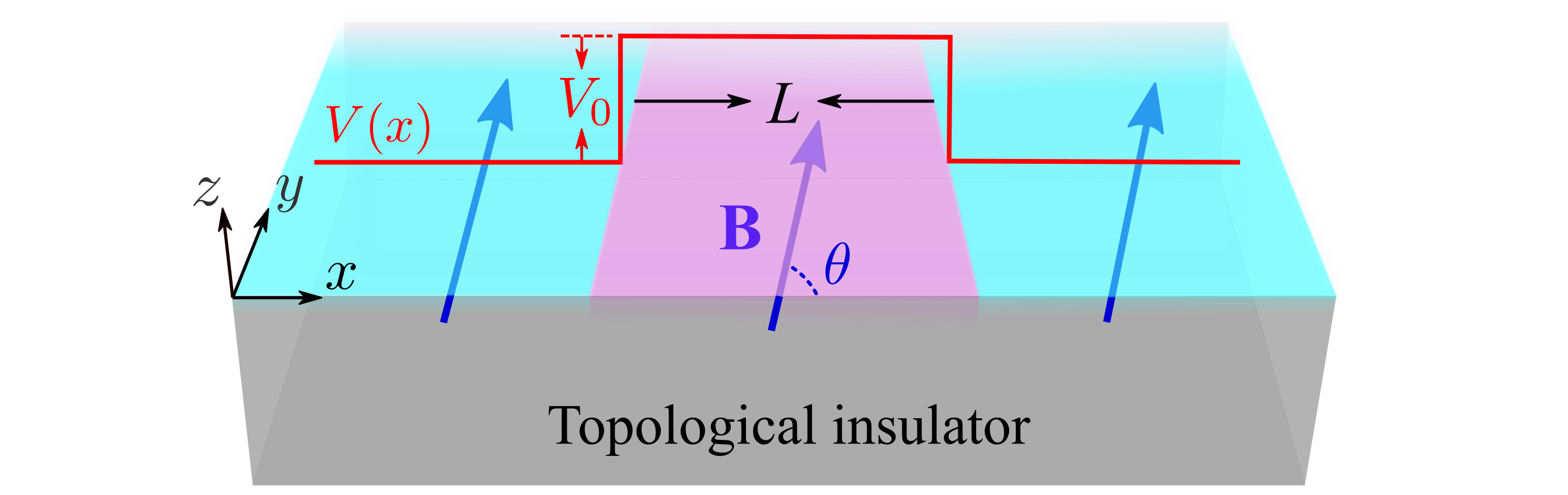}

\caption{Schematic of surface states (cyan and magenta) of a topological insulator (gray) with a barrier potential $V_{0}$ extending over a length of $L$. An in-plane magnetic field ${\bf B}$ (blue arrows) is applied to the system.}

\label{fig:setup}
\end{figure}

In this Letter, we study the conductance of topological surface states in presence of a barrier potential and external in-plane magnetic fields, taking into account the electron-hole asymmetry of the energy spectrum. We find that for small and moderate barrier potentials (comparable to the Fermi energy), the surface states exhibit a positive magnetoconductance due to the increase of the Fermi surface by the tilting in any direction. Remarkably, for larger barrier potentials, super-resonant transport of surface states appear by tuning the magnetic field, which enable many surface propagating modes to tunnel through the barrier without backscattering. This super-resonant transport results in pronounced oscillations in the conductance as strength or direction of the magnetic field are varied. Moreover, we show that for moderate magnetic fields, the conductance is a sinusoidal function of field direction with period $\pi$ and phase shift dependent on the angle between tilting and field directions. Our predictions are applicable to a variety of topological insulators including HgTe and Bi$_{2}$Se$_{3}$ as we argue below.

\textit{Effective Hamiltonian of surface states.}\textemdash The states of a topological insulator on a surface can be described by a single Dirac cone \citep{ZhangHJ_09nphys, Shan11njp,CXLiu10prb}
\begin{equation}
H({\bf k})=m{\bf k}^{2}+v(k_{x}s_{y}-k_{y}s_{x}),
\end{equation}
where ${\bf k}=(k_{x},k_{y})$ are momenta in the vicinity of the $\Gamma$ point, $v$ is the Fermi velocity, $s_{x}$ and $s_{y}$ are Pauli matrices acting on spin space. Moreover, in certain topological insulators, for instance, HgTe with zinc-blende crystal structure, bulk inversion symmetry is broken, leading to extra terms $H_{\text{BIA}}=v_{b}(k_{x}s_{x}+k_{y}s_{y})+\gamma k_{x}k_{y}$ \citep{SuppInf}. Note that we have included the quadratic terms in momentum, $m{\bf k}^{2}$ and $\gamma k_{x}k_{y}$, which preserve time-reversal symmetry. These terms are often ignored in previous studies for simplicity \cite{NoteN6}. However, they break electron-hole symmetry in the energy spectrum and can lead to interesting physics as we show below.

Applying an in-plane magnetic field ${\bf B}=B(\cos\theta,\sin\theta)$ introduces a Zeeman term $H_{\text{Z}}=g\mu_{B}{\bf B}\cdot{\bf s}/2$, where $g$ is the $g$-factor, $\mu_{B}$ is the Bohr magneton, $B$ and $\theta$ denote the strength and direction of the magnetic field, respectively. The Zeeman term not only shifts the Dirac cone away from the $\Gamma$ point in momentum space but also tilts the Dirac cone \citep{SDZheng20prb}. Considering that $v_{b}$ is typically much smaller than $v$ \citep{NoteN2}, we can find the position shift of the Dirac point as ${\bf k}_{s}=k_{s}(-\sin\theta,\cos\theta)$ with $k_{s}=g\mu_{B}B/2v$.  Near the Dirac point, the effective model for surface states can be written as \citep{SuppInf}
\begin{eqnarray}
\mathcal{H}({\bf k}) & = & v(\tilde{k}_{x}s_{y}-\tilde{k}_{y}s_{x})+t_{x}\tilde{k}_{x}+t_{y}\tilde{k}_{y},\label{eq:model}
\end{eqnarray}
where $\tilde{{\bf k}}={\bf k}-{\bf k}_{s}$ and the tilting vector ${\bf t}\equiv(t_{x},t_{y})$ is given by
\begin{equation}
{\bf t}=k_{s}(\gamma\cos\theta-2m\sin\theta,2m\cos\theta-\gamma\sin\theta).\label{eq:tilting-direction}
\end{equation}
The eigen-energies are tilted as $E_{\pm}({\bf k})={\bf t}\cdot\tilde{{\bf k}}\pm v|\tilde{{\bf k}}|.$
The tilting strength $|{\bf t}|$ is proportional to the field strength and the tilting direction is controllable by the field direction. We focus on the realistic case with small tilting $|{\bf t}|<|v|$ throughout.

\textit{Transmission probability.}\textemdash We consider the surface states with a barrier potential $V_{0}$ extending over a length of $L$ in $x$-direction, as sketched in Fig.\ \ref{fig:setup}. The in-plane magnetic field is applied to the whole system. This setup can be described by
\begin{equation}
\mathcal{H}_{\text{tot}}=\mathcal{H}(-i\partial_{{\bf r}})-E_{F}+V(x)
\end{equation}
with $E_{F}$ the Fermi energy and the local electronic potential $V(x)=V_{0}$ for $|x|\leqslant L/2$ and $0$ otherwise \cite{Note1}. $V_{0}$ can be created by local gating \citep{Banerjee19NanoScale}. It can be positive or negative. For simplicity, we assume the system to be large in $y$-direction such that the transverse momentum $k_{y}$ is conserved.

To study the transport properties of the system, we employ the scattering approach. In each region, we find two eigenstates for given energy $E$ and momentum $k_{y}$. In the regions away from the barrier, their wavefunctions can be written as
\begin{align}
\psi_{\pm}(x,y) & =e^{i\tilde{k}_{y}y}e^{i\tilde{k}_{\pm}x}\begin{pmatrix}e^{i\theta_{\pm}},-1\end{pmatrix}^{T}/\mathcal{N}_{\pm},\label{eq:wave-functions}
\end{align}
where $e^{i\theta_{\pm}}\equiv v(\tilde{k}_{y}+i\tilde{k}_{\pm})/(E_{k_{y}}-t_{x}\tilde{k}_{\pm})$,
$E_{k_{y}}=E+E_{F}-t_{y}\tilde{k}_{y}$, $\mathcal{N}_{\pm}=\sqrt{1+|e^{i\theta_{\pm}}|^{2}}$,
and the wavenumbers $\tilde{k}_{\pm}$ in $x$-direction are given by
\begin{align}
\tilde{k}_{\pm} & =[-t_{x}E_{k_{y}}\pm v\sqrt{E_{k_{y}}^{2}-(v^{2}-t_{x}^{2})\tilde{k}_{y}^{2}}]/(v^{2}-t_{x}^{2}).\label{eq:k-formulas}
\end{align}
In the barrier region, the wavefunctions have the same form as Eq.\ (\ref{eq:wave-functions}) but with $E_{k_{y}}$ replaced by $E_{k_{y}}^{B}=E_{k_{y}}-V_{0}$. Correspondingly, we use superscript $B$ to indicate the angles $\theta_{\pm}^{B}$ and wavenumbers $\tilde{k}_{\pm}^{B}$ inside the barrier region.

The scattering state of injecting an electron from the one lead to the junction can be expanded in terms of the basis wavefunctions, Eq.\ (\ref{eq:wave-functions}). Matching the wavefunction of the scattering state at the interfaces, we derive the transmission coefficient as
\begin{eqnarray}
t_{\tilde{k}_{y}} & = & e^{-i\tilde{k}_{+}L}e^{i(\tilde{k}_{-}^{B}+\tilde{k}_{+}^{B})L}(e^{i\theta_{+}}-e^{i\theta_{-}})(e^{i\theta_{+}^{B}}-e^{i\theta_{-}^{B}})/\mathcal{Z},\label{eq:geneal-formula}
\end{eqnarray}
where $\mathcal{Z}=e^{i\tilde{k}_{+}^{B}L}(e^{i\theta_{+}}-e^{i\theta_{+}^{B}})(e^{i\theta_{-}}-e^{i\theta_{-}^{B}})-e^{i\tilde{k}_{-}^{B}L}(e^{i\theta_{+}}-e^{i\theta_{-}^{B}})(e^{i\theta_{-}}-e^{i\theta_{+}^{B}})$. The transmission probability is then given by $T_{\tilde{k}_{y}}=|t_{\tilde{k}_{y}}|^{2}$.  More details of the derivation are presented in the Supplemental Material (SM) \citep{SuppInf}. For the incident modes with $\tilde{k}_{y}=0$, we always have $\theta_{\pm}=\theta_{\pm}^{B}=\pm\pi/2$ and hence $T_{\tilde{k}_{y}=0}=1$. This perfect transmission results from spin conservation and is related to Klein tunneling\ \citep{Katsnelson06nphys}. Notably, without tilting, the results are independent of the magnetic field. This indicates that a simple position shift of the Dirac cone in momentum space does not change the transport properties of surface states.

\begin{figure}[t]
\includegraphics[width=1.03\columnwidth]{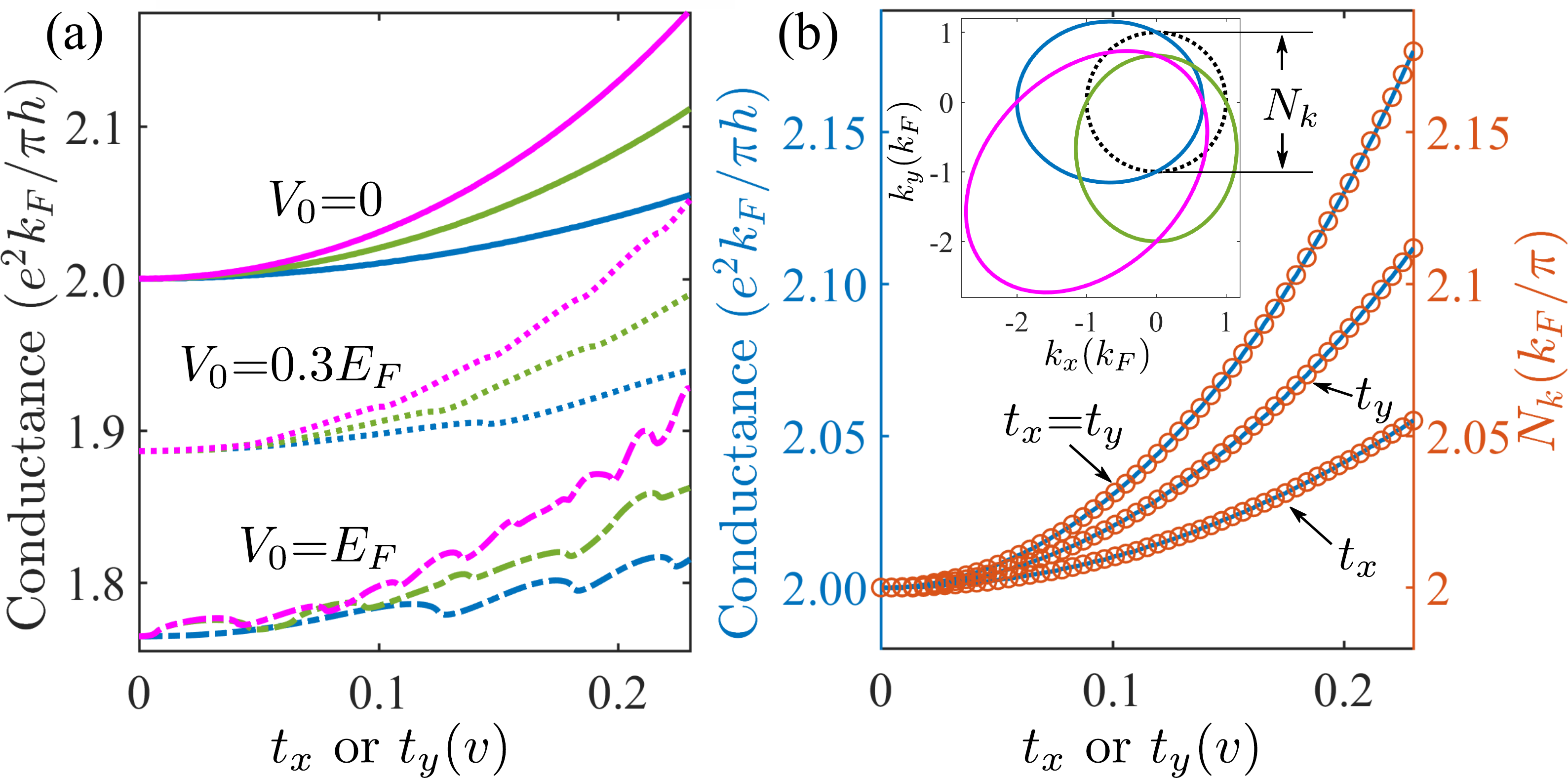}

\caption{(a) Conductance $G$ (in units of $e^{2}k_{F}/\pi h$ with $k_{F}=|E_{F}/v|$) as a function of  tilt strength $t_{x}$ (for $t_{y}=0$, blue), $t_{y}$ (for $t_{x}=0$, green) and $t_{x}=t_{y}=t$ (magenta) for $E_{F}=100v/L$ and  $V_{0}=0$ (solid), $0.3E_{F}$ (dotted), and $E_{F}$ (broken), respectively. (b) Number $N_{k}$ of propagating modes (in units of $k_{F}/\pi$) as a function of $t_{x}$ (for $t_{y}=0$), $t_{y}$ (for $t_{x}=0$), and $t_{x}=t_{y}=t$, respectively. Inset: Fermi surfaces for ${\bf t}=(0.5,0)v$,
$(0,0.5)v$, and $(0.5,0.5)v$.}

\label{fig:positive-MC}
\end{figure}

\begin{figure*}[t]
\includegraphics[width=1\textwidth]{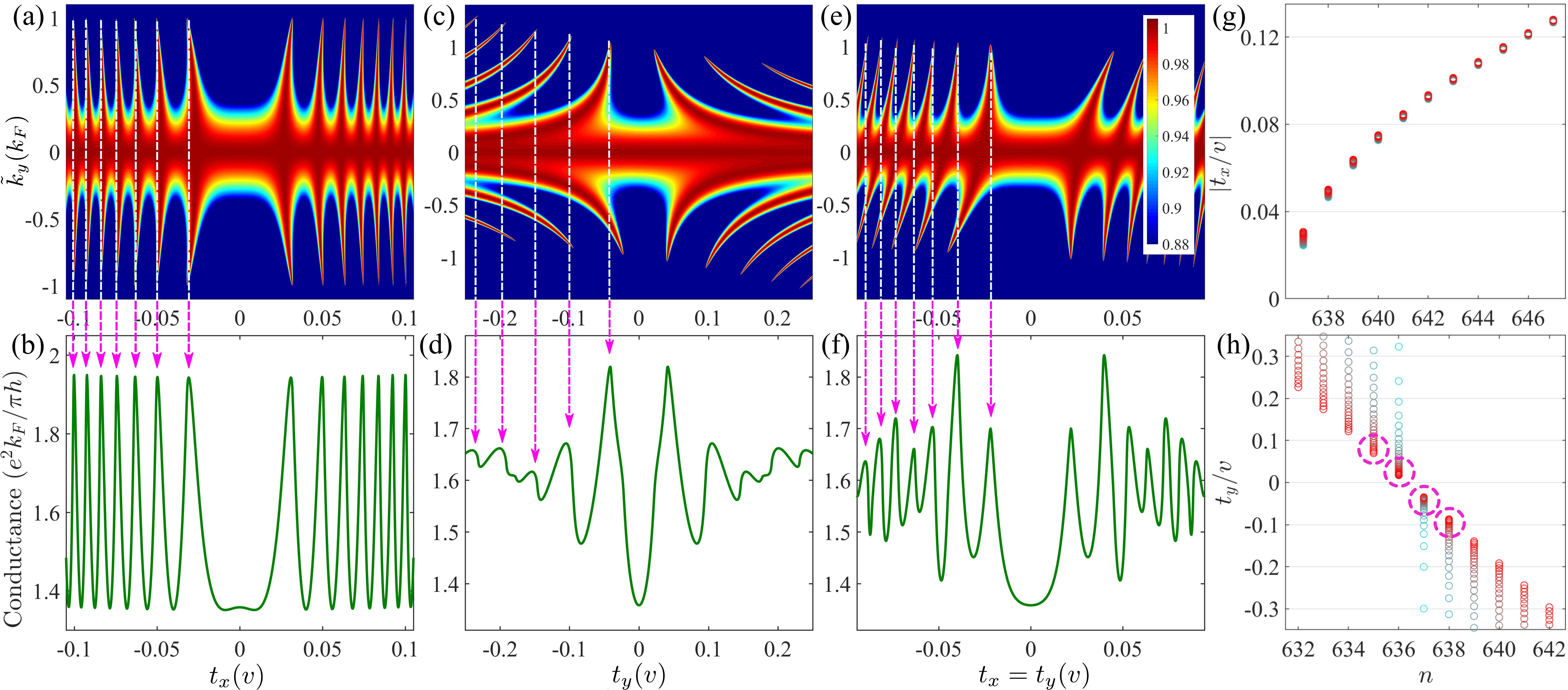}

\caption{(a) Transmission probability density against $\tilde{k}_{y}$ and $t_{x}$ for $t_{y}=0$; (b) Conductance $G$ as a function of $t_{x}$ for $t_{y}=0$; (c) Transmission probability density against $\tilde{k}_{y}$ and $t_{y}$ for $t_{x}=0$; (d) $G$ as a function of $t_{y}$ for $t_{x}=0$. (e) Transmission probability density against $\tilde{k}_{y}$ and $t_{x}=t_{y}=t$; (f) $G$ as a function of $t_{x}=t_{y}=t$. In (b), (d) and (e), the peaks (marked by magenta arrows) of $G$ correspond to the super-resonant regimes. (g) Solutions of $t_x$ to Eq.\ \eqref{eq:resonance-condition} for $t_y=0$ and different integers $n$ and mode indices $\tilde{k}_{y}$. (h) Solutions of $t_y$ to Eq.\ \eqref{eq:resonance-condition} for $t_x=0$ and different $n$ and $\tilde{k}_{y}$. In (g) and (h), the color changes from cyan to red when $\tilde{k}_{y}$ increases from $0$ to $1.2k_{F}$. In (h), the super-resonant regimes are marked by the dashed circles. Other parameters are $E_{F}=vk_F=50v/L$ and $V_{0}=40E_{F}$. We provide more illustrations for other values of $V_0$ in the SM  \citep{SuppInf}.}

\label{fig:oscillations}
\end{figure*}

\textit{Positive magnetoconductance.}\textemdash With the transmission probability, the (differential) conductance $G$ (per unit length) at zero temperature and zero bias voltage can be evaluated as
\begin{align}
G & =\dfrac{e^{2}}{h}\int\dfrac{d\tilde{k}_{y}}{2\pi}T_{\tilde{k}_{y}}(E=0),\label{eq:conductance}
\end{align}
where the sum runs over all modes distinguished by $\tilde{k}_{y}$.

We first look at the case of small and moderate barrier potentials, i.e., $|V_{0}|\lesssim|E_{F}|$, as shown in Fig.\ \ref{fig:positive-MC}. Notably, $G$ increases as we increase the tilting strength in any direction. Recalling that the tilting strength grows linearly with increasing the magnetic field $B$, this indicates a positive magnetoconductance. For small barrier potentials, $|V_{0}|\ll|E_{F}|$, $G$ increases monotonically with increasing $B$. A larger barrier potential suppresses $G$ and induces slight oscillations. However, $G$ increases overall, with increasing $B$ [Fig.\ \ref{fig:positive-MC}(a)]. These oscillations are closely related to the super-resonant transport of tilted surface electrons, which we discuss later.

The positive magnetoconductance can be attributed to the enhanced Fermi surface of tilted surface states. To understand this, it is instructive to consider the zero-barrier limit $V_{0}=0$. In this limit, all propagating modes transmit through the junction without reflection. Thus, the conductance is simply given by the number $N_{k}$ of propagating modes, i.e., $G=(e^{2}/h)N_{k}.$ $N_{k}$ is determined by the size of the Fermi surface in $k_{y}$-direction, as illustrated in the inset of Fig.\ \ref{fig:positive-MC}(b). Tilting the surface Dirac cone in any direction enlarges the Fermi surface and hence the number of propagating modes. As shown by the circles in Fig.\ \ref{fig:positive-MC}(b), we calculate $N_{k}$ numerically as a function of the tilting strength in three different directions as considered in Fig.\ \ref{fig:positive-MC}(a). Evidently, this dependence nicely agrees with the magnetoconductance (solid curves). When the tilting occurs in $x$- or $y$-direction, we can find $N_{k}$ analytically from the tilted spectrum. Namely, $N_{k}=|E_{F}|/(\pi\sqrt{v^{2}-t_{x}^{2}})$ for tilting in $x$-direction and $N_{k}=|vE_{F}|/[\pi(v^{2}-t_{y}^{2})]$ for tilting in $y$-direction.

\textit{Super-resonant transport and conductance oscillations.}\textemdash Now, we consider larger barrier potentials, $|V_{0}|>|E_{F}|$, and analyze the magnetic oscillations of the conductance. These oscillations can be understood as the emergence of super-resonant (transport) regimes of surface states, where many propagating modes perfectly transmit through the barrier at the same magnetic field (i.e. the same tilting). To make this clearer and simplify the analysis, we first focus on the large barrier limit, $|V_{0}|\gg|E_{F}|$. In this limit, we can approximate $\theta_{\pm}^{B}\approx\pm\pi/2$ in Eq.\ (\ref{eq:geneal-formula})
and simplify
\begin{align}
T_{\tilde{k}_{y}} & =\dfrac{1-\cos(\theta_{+}-\theta_{-})}{1-\sin\theta_{+}\sin\theta_{-}-\cos[(\tilde{k}_{+}^{B}-\tilde{k}_{-}^{B})L]\cos\theta_{+}\cos\theta_{-}}.\label{eq:resonance}
\end{align}
From this expression [more generally Eq.\ (\ref{eq:geneal-formula})], we find that the barrier becomes transparent
for the mode with index $\tilde{k}_{y}$ when the resonance condition, $\sin[(\tilde{k}_{+}^{B}-\tilde{k}_{-}^{B})L/2]=0$, is fulfilled. Using the expressions for $\tilde{k}_{\pm}^{B}$ in Eq.\ (\ref{eq:k-formulas}), the resonance condition reads explicitly
\begin{equation}
(V_{0}-t_{y}\tilde{k}_{y})^{2}-(v^{2}-t_{x}^{2})\tilde{k}_{y}^{2}=[n\pi(v^{2}-t_{x}^{2})/(vL)]^{2}\label{eq:resonance-condition}
\end{equation}
with $n$ an integer. This means that an electron acquires a  phase shift $2n\pi$ in one round trip between the interfaces \citep{Noten-physics}.

When the tilting is in junction (i.e. $x$-) direction, we find the solutions of $t_{x}$ to Eq.\ (\ref{eq:resonance-condition}) as
\begin{align}
t_{x} & \approx \pm\sqrt{v^{2}-|vV_{0}|L/\pi n}\label{eq:tx-distance}
\end{align}
for integers $n>|V_0L/\pi v|$. Strikingly, these solutions are independent of the mode index $\tilde{k}_{y}$. This indicates the super-resonant regimes where all propagating modes with different $\tilde{k}_{y}$ exhibit perfect transmission. As a result, we observe resonance peaks in $T_{\tilde{k}_{y}}$ and thus the maximal conductance $G_{\text{max}}=(e^{2}/h)N_{k}$ at $t_{x}$ ($\propto B$) determined by Eq.\ (\ref{eq:tx-distance}). Moreover, we find that at $t_{x}=\pm\sqrt{v^{2}-|vV_{0}|L/[\pi(n+1/2)]}$, all modes have instead the lowest transmission probabilities given by $T_{\tilde{k}_{y}}=1-(v^{2}-t_{x}^{2})\tilde{k}_{y}^{2}/E_{F}^{2}$ \citep{SuppInf}. Summing over all modes, we obtain the minimal conductance as $G_{\text{min}}=(2e^{2}/3h)N_{k}$. Therefore, we observe pronounced oscillations of $G$ with magnitude $\Delta G_{\text{osc}}$ as large as one third of the maximal conductance:
\begin{equation}
\Delta G_{\text{osc}}=G_{\text{max}}/3.
\end{equation}
Interestingly, the values of $G_{\text{max}}$, $G_{\text{min}}$ and $\Delta G_{\text{osc}}$ (in units of $N_{k}$) are universal and independent of the potential $V_{0}$ and length $L$ of the barrier \cite{Note-universal-value}. Considering the increase of $N_{k}$, when strengthening $B$, $\Delta G_{\text{osc}}$ increases. In contrast, according to Eq.\ (\ref{eq:tx-distance}), the separations between the conductance peaks depend strongly on the product $V_{0}L$, whereas they are insensitive to $E_{F}$. Moreover, they decrease with increasing $B$. All these results are in accordance with our numerical results displayed in Fig.\ \ref{fig:oscillations}(a),
(b) and (g).

When the tilting is in transverse (i.e. $y$-) direction, the solutions to Eq.\ (\ref{eq:resonance-condition}) are given by $t_{y}=V_{0}/\tilde{k}_{y}-v\sqrt{1+(\pi n/\tilde{k}_{y}L)^{2}}$. For $n$ close to $n_{c}\equiv[|V_{0}L/\pi v|]$, the greatest integer less than $|V_{0}L/\pi v|$, we find that most of the propagating modes exhibit a resonance condition at
\begin{equation}
t_{y}\approx\pm(|vV_{0}|L-\pi nv^{2})/|E_{F}L|.\label{eq:ty-distance}
\end{equation}
Hence, in this tilting direction, we also observe the resonance peaks and pronounced oscillations in the conductance [Fig.\ \ref{fig:oscillations}(d)]. In contrast to the case with the tilting in junction direction, the separations between the conductance peaks are sensitive not only to $L$ and $V_{0}$ individually but also to $E_{F}$. Moreover, Eq.\ (\ref{eq:ty-distance}) indicates that the separations between the conductance peaks are almost constant with respect to $B$. However, as increasing $B$, the resonance positions for different propagating modes become more extended [Fig.\ \ref{fig:oscillations}(c) and (h)]. Consequently, the magnitude of oscillations is strongest
for small $B$ but suppressed for large $B$.

For the general case with the tilting direction deviating from $x$- and $y$-directions, $t_{x}=\varepsilon t_{y}$ with $\varepsilon\neq0$, we can also observe magnetoconductance oscillations [Fig.\ \ref{fig:oscillations}(e) and (f)]. These oscillations can be similarly attributed to the super-resonant transport of surface states as varying $B$. However, they are less regular, compared to the two special cases discussed above. The oscillations are aperiodic in $B$ and the positions of the peaks become hard to predict in general.

Note that although we focus on the large barrier limit in the above analysis, the conductance oscillations remain pronounced even when the barrier potential is of the same order as the Fermi energy, $|V_{0}|\apprge|E_{F}|$, see Fig.\ \ref{fig:positive-MC}(a) and Sec.\ VI in the SM
\citep{SuppInf}.

\begin{figure}[b]
\includegraphics[width=1\columnwidth]{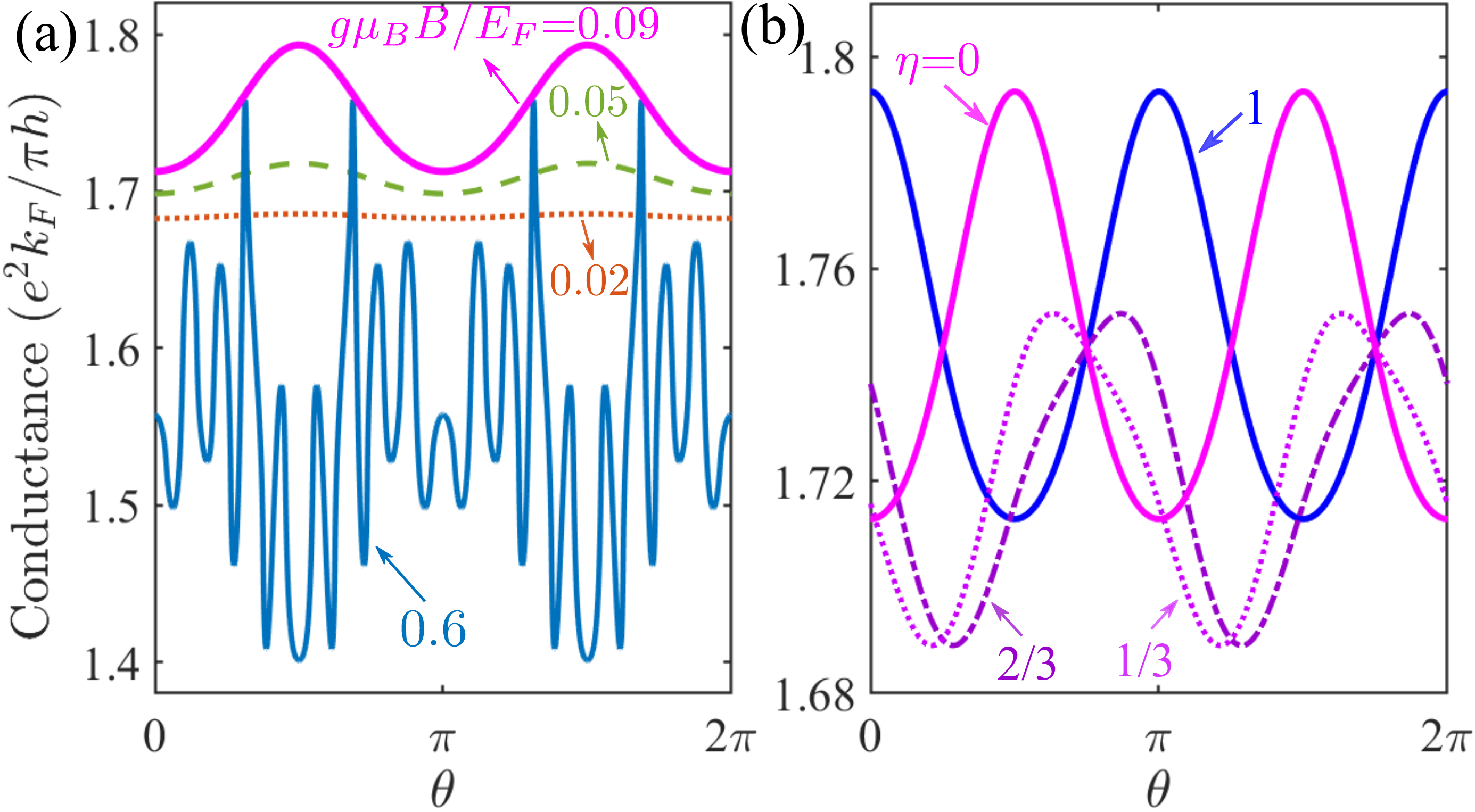}

\caption{(a) Conductance $G$ as a function of field direction $\theta$ for $m=10^{-3}vL$, $\gamma=0$ and $g\mu_B B=0.02E_F$, $0.05E_F$, $0.09E_F$ and $0.6E_F$, respectively. (b) the same as (a) but for $g\mu_B B=0.02E_F$, $m=10^{-3}(1-\eta)vL$ and $\gamma=2\times 10^{-3}\eta vL$ with $\eta=0,$ $1/3$, $2/3$ and $1$ (from magenta to blue), respectively.
Other parameters are $E_{F}=100v/L$ and $V_{0}=40E_{F}$.}

\label{fig:direction}
\end{figure}

\textit{Dependence on field direction.}\textemdash As we have discussed before, the conductance $G$ depends on the tilting direction which, in turn, is determined periodically by the  field direction $\theta$, according to Eq.\ (\ref{eq:tilting-direction}). Therefore,  $G$ depends periodically on $\theta$. This field-direction dependence stems from two origins: (i) the anisotropic Fermi surface and (ii) the barrier transparency for conducting channels. In Fig.\ \ref{fig:direction}, we calculate numerically $G$ as a function of $\theta$ . Several interesting features can be observed.

First, $G$ has a period of $\pi$ in $\theta$. For small field strengths $B<B_{c}$, $G(\theta)$ displays approximately a sinusoidal dependence, $G(\theta)-G_{0}\propto\sin[2(\theta-\theta_{0})]$, where $G_{0}$ is a ($\theta$-independent) constant and $B_{c}$ corresponds to the field strength at which the first conductance peak is located \citep{NoteN}. If the tilting direction is parallel ($m=0$) or perpendicular ($\gamma=0$) to the field direction, the phase shift becomes $\theta_{0}=0$ or $\pi/2$. However, if the tilting direction is neither parallel nor perpendicular to the field direction ($m\gamma\neq0$), then $\theta_{0}$ is different from 0 and $\pi/2$ [Fig.\ \ref{fig:direction}(b)]. Second, if we increase the field strength $B$, the dependence on $\theta$ becomes more pronounced [Fig.\ \ref{fig:direction}(a)]. This shows that the anisotropy of surface states is enhanced by increasing $B$ via the tilting effect. Third, for stronger field strengths $B>B_{c}$, $G$ oscillates with a number of peaks and valleys in one period $\theta\in[0,\text{\ensuremath{\pi}})$ (blue curve). These dense oscillations with respect to $\theta$ can also be related to the super-resonant transport of surface states analyzed before. It is interesting to note that clear field-direction dependence in the resistance of topological surface states has been observed recently \cite{Sulaev15NL,Taskin17nc}.

\textit{Conclusion and discussion.}\textemdash We have identified a positive magnetoconductance of topological surface states, which stems from the increase of Fermi surface by applying in-plane magnetic fields. We have unveiled the super-resonant transport of the surface states by tuning the magnetic field, which enables many propagating modes to transmit a barrier potential without backscattering. This super-resonant transport results in pronounced oscillations in the magnetoconductance.

We note that the appearance of the positive magnetoconductance and conductance oscillations can be directly attributed to the deformation of the surface Dirac cone by in-plane magnetic fields. In this work, the crucial role of deforming is played by tilting the Dirac cone via the Zeeman effect. Particularly, the anomalous conductance oscillations arising from the super-resonant transport of surface states are essentially different from conventional magnetic oscillations, which typically stem from the formation of Landau levels or the Aharonov-Bohm effect.

Our predictions can be implemented in various candidate materials including HgTe and Bi$_{2}$Se$_{3}$ where in-plane magnetic fields have been successfully applied to surface states \citep{He11prl,Wiedmann16prb,Taskin17nc,Sulaev15NL,Hart17nphys,Rakhmilevich18PRB,BWu18APL}.
Consider HgTe with parameters $v=256$ meV$\cdot$nm, $m=108$ meV$\cdot$nm$^2$, $\gamma=-64$ meV$\cdot$nm$^2$, $g=20$ \citep{SuppInf}, $L=2\mu$m, and $V_0=40$ meV \cite{NoteN7}. We could observe magnetic oscillations for tilting $t_{c}>v\sqrt{1-V_0L/[\pi v(n_c+1/2)]}\simeq 0.017v$ and thus for magnetic fields $B> 2t_cv/(g\mu_B\sqrt{4m^2+\gamma^2})\simeq 8.4$ T. For Bi$_{2}$Se$_{3}$ with $v=330$ meV$\cdot$nm, $m=237$ meV$\cdot$nm$^2$, $\gamma=0$, $g=19.4$, $L=2\mu$m, and $V_0=150$ meV \citep{SuppInf}, we could observe the oscillations for $B>8.6$ T \cite{Note-B-value}.

\begin{acknowledgements}We thank Abu Aravindnath, Mohamed AbdelGhany, Wouter Beugeling, Hartmut Buhmann, Charles Gould, and Benedikt Mayer for valuable discussion. This work was supported by the DFG (SPP1666, SFB1170 ``ToCoTronics'', and SFB1143 (project-id 247310070)), the W\"urzburg-Dresden Cluster of Excellence ct.qmat (EXC2147, project-id 390858490), and the Elitenetzwerk Bayern Graduate School on ``Topological Insulators''. P.S. is also supported by the DFG through the Leibniz Program and the National Science Centre Sonata Bis grant 2019/34/E/ST3/00405.
\end{acknowledgements}


\begin{thebibliography}{70}
\bibitem [{Not({\natexlab{a}})}]{NoteN2}%
  \BibitemOpen
 \bibfield  {journal} {  }\bibfield  {title} { {\bibinfo {title} {We
  note, however, that the main results discussed below are not restricted to
  this condition}}}\href@noop {} {\ }\BibitemShut {NoStop}%
\bibitem [{Not({\natexlab{a}})}]{NoteN6}%
  \BibitemOpen
 \bibfield  {journal} {  }\bibfield  {title} { {\bibinfo {title} {In a more elaborate model, we may include the warping effect of surface states. However, using realistic parameters for Bi-based topological insulators, we find that the influence of the warping effect on our main results is ignorable \cite{SuppInf}}}}\href@noop {} {\ }\BibitemShut {NoStop}%
\bibitem [{Not({\natexlab{b}})}]{Noten-physics}%
  \BibitemOpen
  \bibfield  {title} { {\bibinfo {title} {We may look at the system as a Fabry-P\'eroit resonator between two interfaces. An electron can circulate in the resonator. In one round trip, the electron acquires a phase shift $(\tilde{k}_+^B-\tilde{k}_-^B)L$}}}\href@noop {}\BibitemShut
  {NoStop}%
\bibitem [{Not({\natexlab{b}})}]{Note-universal-value}%
  \BibitemOpen
  \bibfield  {title} { {\bibinfo {title} Note that this is only true when $|V_0|\gg |E_F|$. When this condition is softened, the value $ \Delta G_{\text{osc}}/G_{\text{max}}$ becomes smaller}}\href@noop {}\BibitemShut
  {NoStop}%
\bibitem [{Not({\natexlab{a}})}]{NoteN7}%
  \BibitemOpen
 \bibfield  {journal} {  }\bibfield  {title} { {\bibinfo {title} In HgTe, the phase coherence length can be larger than 5 $\mu$m at temperature $T=40$ mK \cite{Ziegler18prb}, and the bulk energy gap can be larger than 50 meV \cite{Wu14EPL,Rauch15PRL}. To the best of our knowledge, the in-plane g-factor of topological surface states in HgTe is still not known experimentally. However, we notice that in a related 2D system, the quantum spin Hall phase of HgTe quantum wells, the 2D electron gas has an in-plane g-factor of 20.5 \cite{Konig08JPSJ}. We may expect a similar g-factor for the surface states. }}\href@noop {} {\ }\BibitemShut {NoStop}%
\end{thebibliography}

%

\end{document}